# Extracting and visualizing a new classification system for Colombia's National Administrative Department of Statistics. A visual analytics framework case study

Pierre Raimbaud[1,2][0000-0002-5584-8100], Jaime Camilo Espitia Castillo[1], John Alexis Guerra Gomez[1][0000-0001-7943-0000]

[1] Systems and Computing Engineering, Imagine Group, Universidad de los Andes, Bogota, D.C. , Colombia
`p.raimbaud@uniandes.edu.co`
[2] LiSPEN, Arts et Métiers, Institut Image, Chalon-sur-Saone, France
`pierre.raimbaud@ensam.eu`

**Abstract.** In a world filled with data, it is expected for a nation to take decisions informed by data. However, countries need to first collect and publish such data in a way meaningful for both citizens and policy makers. A good thematic classification could be instrumental in helping users navigate and find the right resources on a rich data repository as the one collected by Colombia's National Administrative Department of Statistics (DANE). The Visual Analytics Framework is a methodology for conducting visual analysis developed by T. Munzner et al. [T. Munzner, Visualization Analysis and Design, A K Peters Visualization Series, 1, 2014] that could help with this task. This paper presents a case study applying such framework conducted to help the DANE better visualize their data repository, and present a more understandable classification of it. It describes three main analysis tasks identified, the proposed solutions and the collection of insights generated from them.

## 1   Introduction

The DANE (National Administrative Department of Statistics) is one of the most relevant organizations regarding data in Colombia. It is responsible for planning, collecting, analyzing and distributing the country's national statistics. The total amount of data that this public institution owns is one of the largest in the country (among government institutions), due to the fact that it periodically gathers information about all the major topics of the country. Its information ranges from population statistics, passing by technological literacy, to public access to services, among many others. Because of this, one of DANE's main goals is that public policies in Colombia become more data-driven [2]. However, this is rarely the case, as public institutions sometimes cannot access the information they need because it is not publicly available, or not easy to access, or not well classified and organized. Aware of this, the DANE is seeking to improve the availability and organization of the data they release. With this in mind, representatives from the organization reach out to our university for help applying visual analytics methods to deliver better tools to the different stakeholders of public policy-making structures.

Concretely, the DANE owns data coming from both statistical operations and from administrative records. This paper addresses two main types of data: first the administrative records (called administrative registers afterwards) collected by the DANE, and second the derived statistical analyses conducted on them (called statistical operations afterwards). Figure 1 illustrates this main distinction.

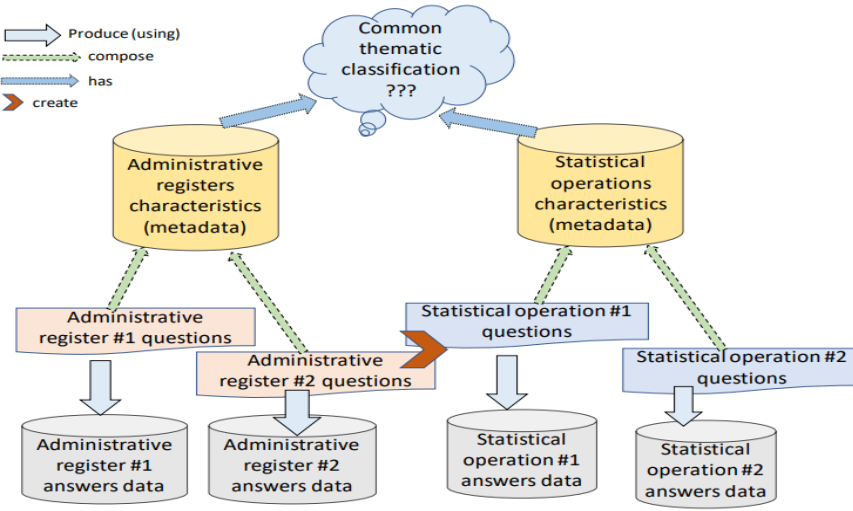

**Fig. 1.** DANE's data distribution and explanation of the problematic: does a new thematic classification coming from the metadata of the administrative registers and statistic operation exist?

As a crucial point for understanding our approach, note that, in this paper, we will not consider the data collected by the DANE when they apply the questions or the requests contained in an administrative register or a statistical operation. We will consider the characteristics of the administrative registers and the statistical operations themselves, meaning their attributes, the fields that they address, in other terms all the metadata that define them. So, our original data will be inventories of statistical operations and administrative registers held by the DANE.

Based on these considerations, the main objective of this project is to build a tool to understand and visualize the topics and keywords present among different groups of statistical operations and administrative registers, ultimately allowing decision-makers to have right overviews of topics and to find which statistical operations and administrative registers are related to one in particular (see Figure 1).

## 2    Related work

### 2.1    Tamara Munzner's framework

For this project, we used Munzner's visualization framework [1] to abstract and understand the data, the user's tasks and to choose the best idioms that allow the users to complete these tasks. It has three dimensions: the WHAT, the WHY and the HOW.

**WHAT**: It refers to the available information (data) for the visualization. The basic abstractions of the dataset arrangements are tables, networks, fields and geometry. In a dataset, we can find items, or nodes and links, and its attributes, etc. Moreover, data can be static (there are no new data over time) or data can be dynamic (typically a data stream) and finally, the items/nodes attributes can be ordered or categorical.

**WHY**: It refers to the tasks abstraction that must feature mainly one action (a verb) and one target (a noun). Here the main objective is to clarify what is the main purpose of a visualization, and its potential secondary purposes. Task abstractions can vary from high to low level (meaning depending on how precise you want to define it), and range from presenting trends (at high level, it would be to consume data, in opposition to produce data) to identifying outliers (here it is already defined at low level). .

**HOW**: It refers to the design decisions taken to visualize the data and to perform the required tasks (meaning both visual representations and interactions if there are some). The two objectives here are to decide which visual channels like size, color, etc. will represent the data, and to choose the right marks, or the visual representations for the data (geometric primitives like lines, points, areas) for the visualization. In this stage, the idea is to choose the visual encoding and the idiom (or representation) that best suits the WHAT and WHY, to finally develop the visualization accordingly.

To illustrate this concept, we want to present very shortly some examples of visualizations that we could use further in this work. First, a bar chart (HOW) allows to summarize distribution (WHY), and to show extremes (WHY) if it also uses order (ascending or descending). Indeed, Elzer et al. [3] showed its efficiency for this kind of tasks, but note that another possible idiom for these tasks is the stacked bar chart, as Indramoto et al. [4] explained it. However, in the stack bar chart, the focus is more on combining single-attribute and overall-attribute comparisons rather than making only single-attribute comparisons for one or more dataset (this is our case, see section 3). Furthermore, notice that here we derived the original dataset, a table, to a network dataset. In this case, following Munzner's framework, this kind of dataset is composed by nodes and links (whereas tables are composed by items) - note that it can be relevant to show these links or not, depending on the task. And about our project data, remember that one of our aims is also to discover a new thematic classification. As Ochs et al. [5] showed it, ontologies manipulations and representations are crucial nowadays, but required much work, so they presented a software framework for doing these tasks: derivation, clustering and visualization as a network. Based on their study, we can note that another visualization for ontologies is the treemap [6]. In our case, we used this last representation, and the radial force representation (see section 3), and note that in both cases, one of the most critical point is the usage of forces in order to separate the nodes, depending on one attribute or relationship. Hilbert et al. explained the usefulness and importance of the forces in a network visualization ; indeed, forces allow to separate and form groups, also called clusters [7]. This approach is useful for our work because we want to permit the public policy makers to make decisions based on visualizations that show a new classification, so in this case it could be shown thanks to the use of clustering (see Figure 2).

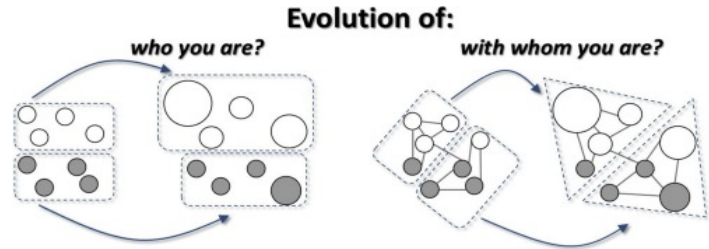

**Fig. 2**: Network visualization with forces for clustering.

## 2.2 Projects with similar issues

In this section, we would like to present some related work that faced the same issues that we are facing in this project, either from the point of view of the policy makers or from the point of view of the designers of visual analytics tools.

First, about public policy and data-driven policy making, Petrini et al. [8] were facing a issue similar to the one of the DANE: Brazil had useful data about some activities in their cities, but the authorities were not using them for public-policy making, even more precisely these data were not used for prioritizing the different public policies over the country. So they applied an analytic hierarchy process (AHP) that used their data and allowed them to provide some visualizations to the policymakers. As there were evaluating various kind of priorities at the same time (environmental, economic, social), they used a stack bar chart for their visualization as shown in Figure 3.

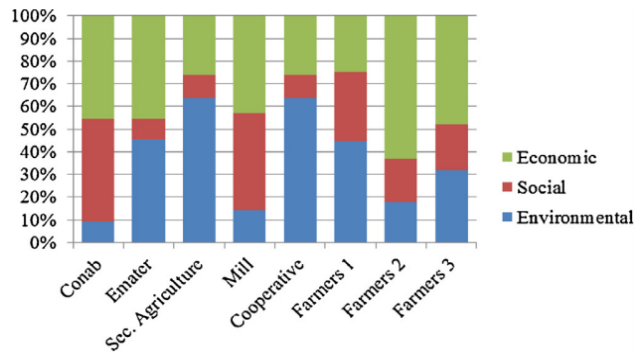

**Fig. 3** Classification of thematic by priorities (multiple values for the priorities).

But for those purposes, first we need some clean data and metadata: it is a common but complex issue to be dealing with unclean data. As Liu et al. [9] showed it, this is a compulsory phase for creating visualizations. In the same paper, they proposed a framework for cleansing and then creating visualizations, as shown in Figure 4.

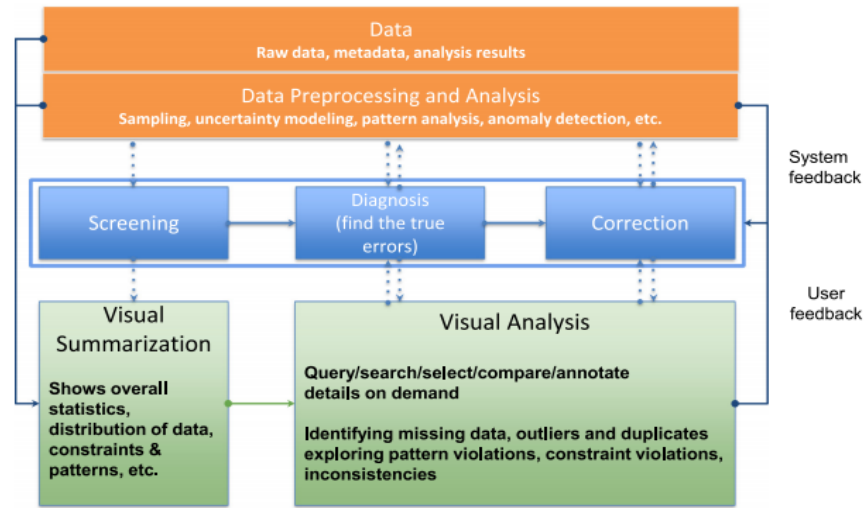

**Fig. 4** Visual analytics framework for steering data quality

We can note that their process could be a complementary approach to Munzner's one, because they gave more importance to the steps of creation and evaluation of the visualization, whereas the Munzner's framework has more focus on the abstraction of the information related to the visualization (data/task/idiom i.e. what, why, how).

Moreover, even when some ontologies have been created especially for the policy makers or other final users, there are real needs of availability and accuracy, meaning that, otherwise, these ontologies would be useless and so not be used by the final users. About this issue, Kamdar et al. made a study about the usage/the access of the users to the ontologies in the biomedical field [10], which queries made the specialists, and eventually how they combined the results of various ontologies. So, in this paper, we can see the importance of creating and then owning ontologies, and the fact they must be user-designed or task/issue-designed and that they can also be viewed from a "macro" point of view, where ontologies can be combined between themselves. In the Figure 5, we can see how can ontologies be built.

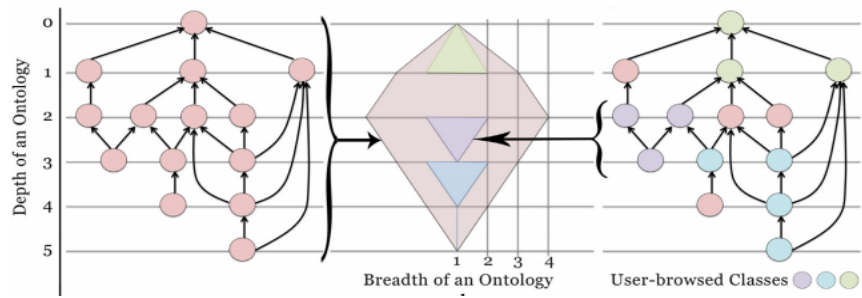

**Fig. 5** Analysis of the construction/composition of ontologies and their depth

To sum up, in this section we have seen some common issues with our project: ontologies or classifications are truly required for policy making, sometimes with an additional classification (priority: "meta-classification", Petrini et al. ). Then we have noted that other frameworks for creating visualizations than the Munzner's one exist, with other focuses than abstraction, cleansing for example (Liu et al. ). We have also seen that these ontologies or classifications for being useful need to have a good accessibility and accuracy (Kamdar et al. ). That leads us to our case study: the classification of the metadata of the DANE data and its usage. Currently, the DANE public policy-making tools don't satisfy the final users of their tools (policymakers), because the classification used does not fit with policy making, and the visualizations are not appropriate to the tasks that the policymakers want to perform. Particularly, they need to be able to discover (identify) easily which statistical operation or administrative register is the closest and more useful for a specific policy. Figures 6 and 7 show examples of the current classification and visualizations. The classification may be too generic, or specific but not with the right terms, not the ones that really define the information contained in the registers and the operations, and the visualizations don't show where this information is (which surveys/census are the most relevant).

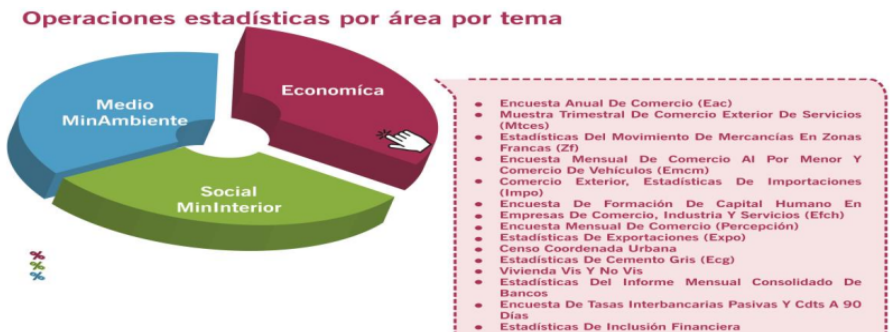

**Fig. 6** : Classification by global/macro topic (source: DANE website)

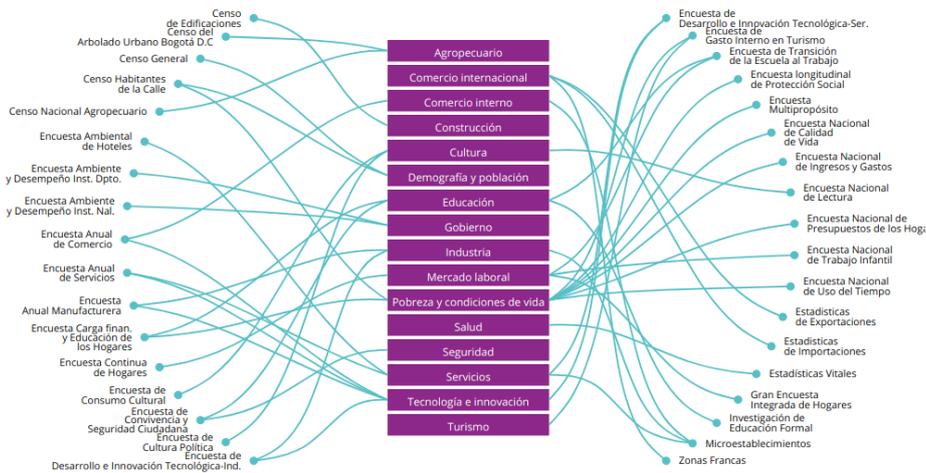

**Fig. 7** Classification by sub themes and link to surveys (source: DANE website)

## 3      Applying the visual analytics framework

In this section, we present how we applied the framework for our three main tasks. But, before explaining each task and the work done for each one, we would like to explain shortly how did we built a new classification using the available metadata, resulting in a new dataset, derived from the previous ones, so we can build the visualization for T2 and T3 using it. Actually, applying the visual analytics framework, it can be considered as task **T0**, called derivation task (from a dataset to new one). For this purpose, we decided to use natural language processing on our original datasets. It was an acceptable solution: the datasets were about 500 lines and 20 columns maximum after cleansing (three CSV files). In our approach, we wrote our own natural language processing tool, but it should be also correct using an existing tool. We decided to create our own tool because we wanted to process the words and build the new dataset in the same program. The process was the following: read all the lines of the file, build a dictionary of keywords that are repeated more than X times in the files (using an exclusion list for the "obvious" words such as determinants or some specific words such as register), and build the nodes and the links of the new dataset, based on the occurrences of the keywords in the metadata of each item of the original dataset.

Then, the three main tasks in this case study were:

**T1**: How many statistical operations and administrative registers are there for each topic (in the original classification) ?

**T2**: How many statistical operations and administrative registers are there for each new topic, considering a new classification coming from the metadata ?

**T3**: Which administrative registers and statistical operations are more related to a specific topic (new classification) ?

For each of these tasks we are going to apply the framework by presenting the WHAT, WHY and HOW abstractions, and our prototypes. Here you will find a summary for each task, then in the next subsections we give more details about each one.

**Task 1 (T1)**

*What*: the original datasets - three tables, two of administrative registers and one of statistical operations (items) with, among others, the following attributes: name (categorical attribute) and thematic area (categorical attribute)

*Why*: **summarize the distribution (considering the old classification)**

*How*: idiom: bar chart ; mark: lines (color hue between the two bar charts, one for the registers, one for the operations) ; channel: vertical position

*Prototype*: see Figures 8 and 9

*Principal insight* (see more in section 6): large difference between the numbers of administrative registers and statistical operations on the macro-category "Economics"

**Task 2 (T2)**

*What*: a new dataset derived from the previous tables - a network where the nodes (items) are the administrative registers, or the statistical operations, or the keywords of a new classification, and the links represent when an administrative register or a statistical operation matches with a keyword, one or more time; some attributes: name (categorical attribute) and new keyword groups (categorical attribute)

*Why*: **summarize the distribution (considering the new classification)**
*How*: idiom: treemap ; mark: point ; channel: color hue, spatial region
*Prototype*: see figure 10 (+11 as an auxiliary visualization for details on demand)
*Principal insight* (see more in section 6): "Labor market" was the penultimate sub-theme in the old classification vs. "Companies" is the 4th with our new classification

**Task 3 (T3)**
*What*: a new dataset derived from the original ones (the same as in task 2)
*Why*: **identify features/extremes** (which nodes are **more related** to a theme, **with the new classification**)
*How*: idiom: radial force ; mark: points ; channel: radial position and color hue
*Prototype*: see Figure 12
*Principal insight* (see more in section 6): choosing the keyword "Health", the most important administrative register is "Individual register of health service delivery–RIPS", and so on with other keywords.

As explained in previous sections, the two main objectives in this paper are: first, to determine new topics (they must be useful for decision making on public policies) that can emerge from the metadata of the administrative registers and the statistical operations, and to evaluate which different statistical operations and administrative registers are more linked to a topic, and secondly, once found this information, to be useful for the policymakers, to provide some appropriate visualizations, typically through a visual analytics tool. Therefore, we developed for this case study a visual analytics tool applying the framework explained above. As there were three main different tasks, it is composed of three main components: first, for the task T1, several context visualizations to analyze the current state of the information held by the DANE (3.1), then for the task T2, a treemap visualization to understand the results of the natural language processing used to understand better the major topics (a new thematic classification useful for decision making) around the DANE's datasets (3.2), and finally for the task T3, a radial force visualization to navigate between and into the identified topics and provide a final tool for policy makers (3.3).

### 3.1 Task 1: general and contextualization task, on the original dataset

The first set of visualizations aims to represent the current inventory of statistical operations and administrative registers held by the DANE. As a result, the main task T1 (WHY) is to *summarize* the *distributions* of both datasets according to different criteria, to answer the following questions:

- How many statistical operations and administrative registers does the DANE have?
- What is the proportion of statistical operations and administrative registers in the three major topics (economics, social and environmental)?
- What is the proportion of statistical operations and administrative registers in each of the 30+ specific topics (for example, health, education, etc. )?

The datasets that we used were two inventories of administrative registers and one inventory of statistical operations provided by the DANE (WHAT). The inventories dataset type was a table for the three datasets.

Based on the analysis made using the Munzner's framework, and as explained before that, the best visual encoding (HOW) to provide this overview is to use some bar charts where the statistical operations and administrative registers are differentiated by colors. With this chosen encoding, it is easy to provide the user the summary of the DANE's information inventories. The Figure 9 shows the first three general and context visualizations developed, and the Figure 9 shows two other visualizations developed to present the distribution of the attribute "sub-theme", allowing to know the global distribution of these sub-themes for all these administrative registers and statistical operations. Note that here we used the same encoding because it is still the best one for *summarizing the distributions* and also to *identify extremes* (secondary task - as explained before that, here we additionally used the technique "separate order and align" for that purpose). So, about identifying extremes, if considering only the administrative registers (in blue, left), the most present sub-theme is "currency" whereas if considering only the statistical operations (in orange, right), it is "agriculture".

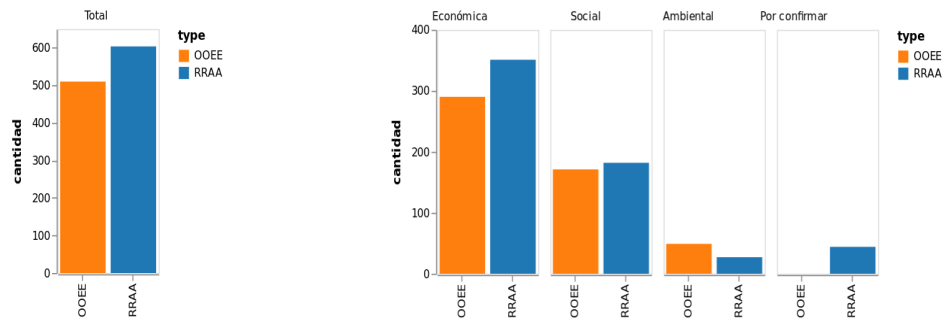

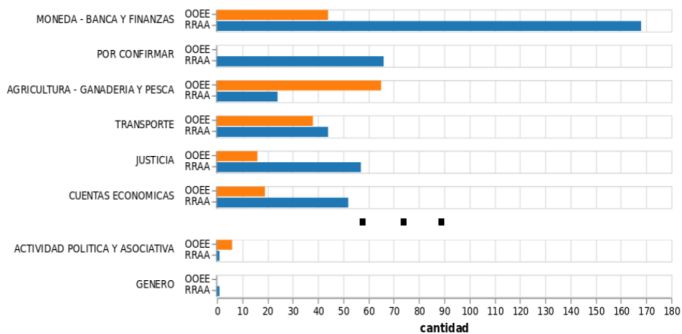

**Fig. 8** Bar chart visualizations based on sub-themes from the original data.

**Subtemas - ordenados**

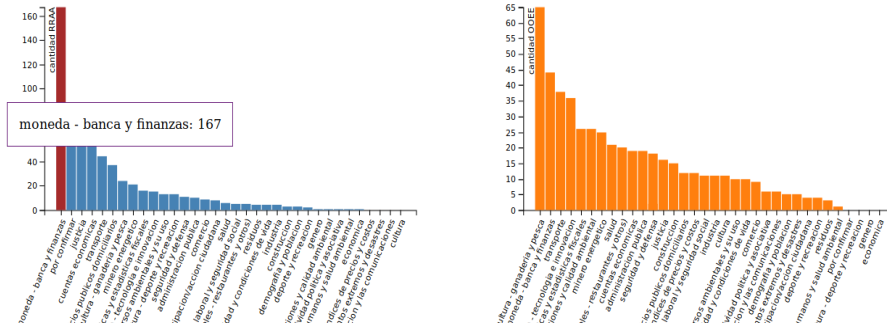

**Fig. 9** Bar chart visualizations about original sub-themes, with "separate order and align".

### 3.2 Task 2: new dataset, new classification, but which distribution ?

Here we want to present the visualization that we have made for the following task T2: *summarize the distribution with the new classification* (WHY). Here we used a treemap (HOW). This treemap uses the derived dataset, from T0 task, that contains nodes and links (WHAT), each node being a statistical operation or an administrative register, and each link being a relationship between two nodes, here in particular between a "register/operation" node and a "keyword" node. We used clustering here for grouping them by (new) theme. For separating the clusters, we used an algorithm called force-in-a-box, made by J. Guerra (https://github.com/john-guerra/forceIn-ABox). As a result, this visualization allows to have a global vision of the new different themes: "transport", "research" etc. (*summarize distribution*). It also allows to detect that "services" and "credits" are the themes that most appear (*identify extremes*).

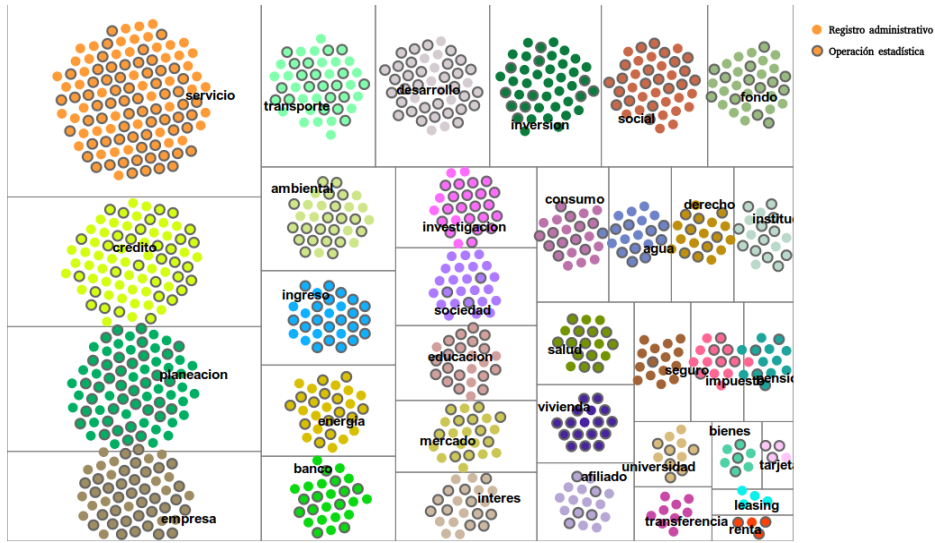

**Fig. 10** Tree map chart visualization on new themes (derived data).

The following visualization is a table coupled to the previous treemap that give information about one item by clicking on it in the treemap,to get specific information .

**Fig. 11** Auxiliary view, a table, of the treemap visualization.

### 3.3 Task 3: given a keyword, which are the items more linked to it?

For navigating between the new generated topics and the nodes associated to it, we created a visualization where the user can type a word and then explore the statistical operations and administrative registers that feature this keyword in their metadata. As a result, the main task T3 of the visualization is to *identify features* (WHY).

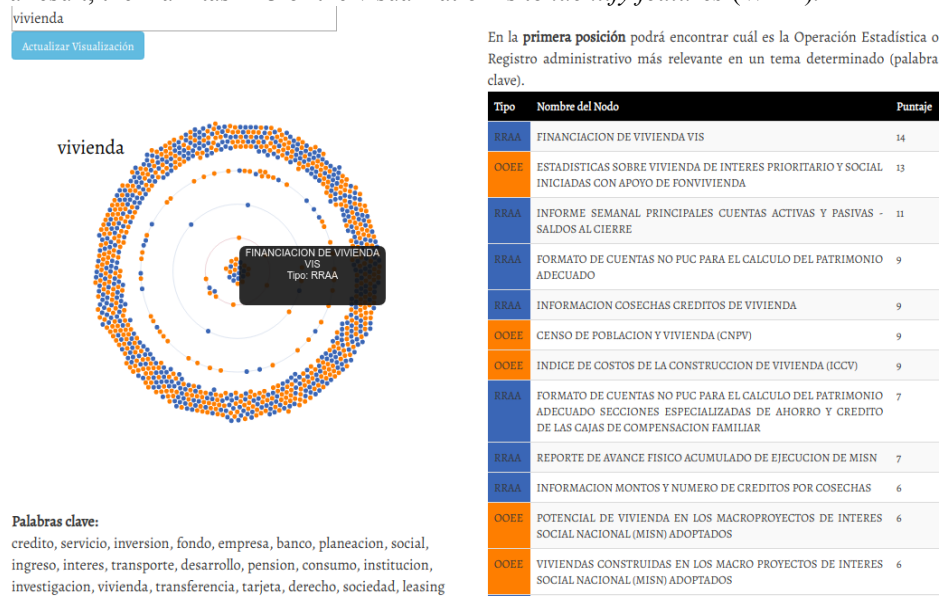

**Fig. 12** Radial force visualization and its auxiliary view, a table, on the right.

In this radial force visualization, after choosing the keyword, the user can see that the statistical operations and administrative registers that contain the keyword are more or less attracted to the center depending on the number of coincidences (the ones that do no contain the keyword keep in the border), allowing to identify the extremes. Thus, in other terms, leaving in the nucleus the most related items, and leaving the less related towards the outskirts of the radial visualization. The Figure 12 shows the radial force visualization, where the statistical operations are orange and the administrative registers are blue. In addition, if the user puts the mouse over any item, he/she will see the item name and its type.

Moreover, to help *identifying the extremes* (WHY), right to the visualization, we added an auxiliary view, a table, where the elements are ordered in descendant order, so the user can find the statistical operation or the administrative register that features the major occurrences of the keyword, considering all of its attributes, in its metadata.

## 4    Experiment and results

### 4.1    Experiment

To validate our work, we organized an experiment where the experts from the DANE were invited to try our tool with all the visualizations created according to the tasks explained before, and according to our application of the visual analytics framework . In total, there were 8 participants , 6 females and 2 males. 5 of them were working in the R&D department (in other words, "our clients", the people who asked for the tool) and 3 were working in the department responsible for the planning based on statistics (in other words, the final users of our future tool, apart from the policymakers). All the participants had to follow a story board that we provided to them first. Here we give a short summary of it: "first, try to get new themes about registers and operations (with the treemap – actually, during the experiment a network visualization, see section 4.2) ; then get some information about one item (with the coupled table); after that, write a word about one theme of your interest and discover which are the registers and operations with more relation with this keyword (with the radial visualization) ; finally read the most important register or operation in the coupled table".

### 4.2    Results

First, note that in this experience, the users were using and judging a previous version of our visual analytics tool that the one presented in this paper. Actually, it is thanks to the results of this experience that we have been able to modify/correct our visualizations to allow the users to get more insights with the visualizations. Nonetheless, these results are interesting because, first, it shows how our work have evolved and moreover it shows that to apply the visual analytics framework may often require to be an iterative process with users experiments (and that some tasks may require more iterations than others). The following results come from a questionnaire that the users filled after the experiment, where they evaluated the quality of our visualizations (usability and completion of the tasks). That is why we asked closed questions using Likert scale, one for each visualization and task. Moreover, as it was an expert (experts on the data used) evaluation, we asked open questions to get some feedback about the visualizations, to make corrections. Finally, according to the results, that would be discussed in the next section, we created the final visualizations that are presented in the paper in section 3. So notice that in this experience, the tool had only two parts not three (the context visualizations part was missing), composed of four visualizations: a network visualization (where the forces were not separating the clusters of nodes as good as in the treemap, which is the "evolution" of this visualization after the experience) and its coupled table, and the radial force visualization and its coupled table.

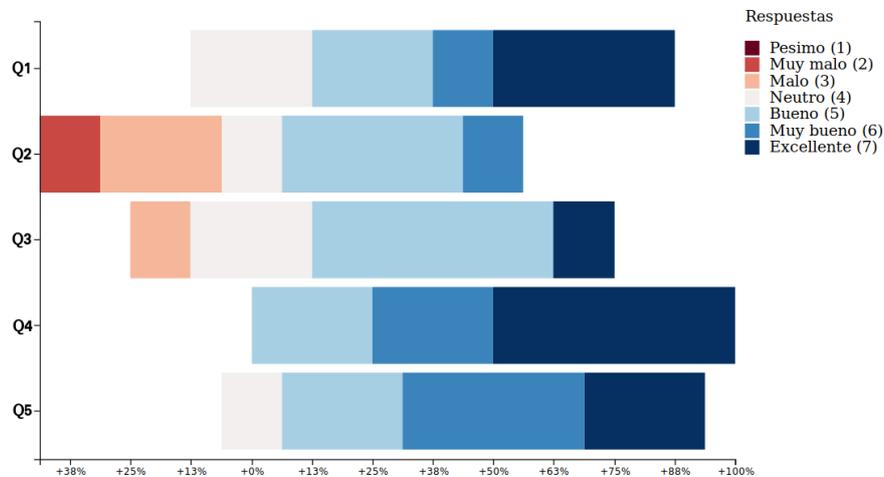

**Fig. 13** Results of the experiment with the users (Likert scale graph).

- Q1- What is your general impression?
- Q2- Have you been able to explore a new classification of the items?
- Q3- Have you been able to obtain the detail for one of these items?
- Q4- Have you been able to discover the items in relation with a theme?
- Q5- Have you been able to identify the item more related with a theme?

## 5  Discussion

According to the results, both the quality results shown on the Likert scale graph (see Figure 13) and the feedback given by the experts indicated that the task they performed with more difficulty (almost 50% of the participants grade it between 1 and 4 included) was the one asked in Q2: explore the new classification in the network visualization. On the contrary, the easiest task for them (100% of them grade it between 4 and 7 included) was the one asked in Q4: discover the items in relation with one theme in the radial force visualization. As a result, we did more corrections on the visualization used in Q2, so it is the only visualization where we had to completely modify the idiom (visual encoding) used, transforming the visualization from a network visualization (clusters might not be appearing so clearly, but the relationships/ links between elements do appear better) to a treemap (the focus is clearly on clustering) – to sum up, both use nodes and forces, but the clustering was clearer in a treemap than in a network visualization, at least in our case.

Additionally, we noticed, both in the comments and the qualitative results (apart from Q2 results, in particular thanks to Q1 results, with 25% of neutral grade - 4), that something might be missing, apart from the current visualizations and their corrections), something that shows better the purpose of our tool and why did we create it.

In other terms: to understand where we are going, we should know what were the original data ? Were there some insights in the original data ? The next visualizations are showing different data ? What is the difference with the new thematic classification? What are the new insights ? All these questions can be considered as a first task for the users. So, that is why we finally added the context visualizations part and organized the tool into three parts and not two: context (original data), treemap (derived data) and radial force visualization (derived data).

## 6   Insights

The new (corrected) visualizations (presented in the section 3) allow to discover some insights about the statistical data and their classification in new themes.

First, thanks to the context visualizations based on the initial data, we can discover than globally more information could be generated because there are more registers than operations (remember that the administrative registers produce and create the statistical operations). More precisely we notice this particularly for the macro category "Economics". Looking at the visualization by sub-theme, the previous insight is confirmed, and we can observe it with more details: the sub-themes "Currency, banks and finance" and "Accounts and economics" appear as some of the ones where there is much difference between the number of registers and the number of operations, and both are belonging to the macro category "Economics".

Then, thanks to the new derived data and the treemap visualization, we can discover other insights. First, by looking to the terms that appear, on one hand this visualization confirms that the macro category and sub-theme currently used are quite coherent with the metadata. For example, we can found as new keywords "research", "credits", "market", whereas that, in the current sub-theme, there are "education, science, technology and innovation", "commerce", "accounts and economics" etc. But, in the other hand, this visualization also suggests that the categories and sub-themes used currently may not be so accurate in term of proportions, because in the current classification, "labor market" is the penultimate sub-theme whereas "companies" is the fourth one with our new thematic classification. Another insight is that this visualization gives us more details about a previous insight from the first visualization component: in the current category "Economics", the focus for generating new operations from registers should be done more precisely on the ones that contains in their metadata the words "leasing" or "transactions" because we can observe in this visualization that there are no operations about these subjects, only registers.

Finally, some other insights have been revealed thanks to the radial force visualization. As our final tool in this study for decision making on public policies, we can notice that thanks to it, people, without being an expert about data manipulation, can identify easily which registers and operations are more related to a theme: for example about "Housing" (the chosen keyword), the most important administrative register is "Housing financing VIS", or with the keyword "Health", the most important administrative register is "Individual register of health service delivery – RIPS". To conclude, this last visualization, by grouping in the center the elements with more rele-vance but separating them into two categories, for being administrative registers or statistical operations by using two different colors, allows at a glance to notice for one thematic if the relevant elements are mostly of one kind of information (because visually it will be mostly of one color), and as a result it can confirm that more statistical operations should be generated in this thematic or not (so we could make some "priorities" about statistical operation generations). So, for example, following with the common theme "economics", we can notice for the keyword "credits" that more statistical operations could be generated (there are more administrative registers).

## 7  Conclusion

Finally, thanks to our approach, we confirmed that the DANE owns highly relevant information for the country and that they should continue in that way, with its efforts to develop more data analysis tools, to provide its different stakeholders with tools for maximizing the usage of the data. In particular, visual analytics tools permit both policymakers and citizens to locate where the information is, and allows its understanding, ultimately enhancing policies and fostering data-driven businesses. So we might think that our contribution helped the DANE to understand that the data they held are very valuable, and that with such approaches, these data can be classified and presented in a way that allow the policymakers to use it for public policies making.

About future work, it could be interesting to explore other possibilities about our natural language processing tool for catching the keywords that appear in the metadata and how could we grade differently the administrative registers or statistical operations that belong to a theme (because, currently it is only based on the number of occurrences of the keywords in the metadata – the DANE has already confirmed it interest for this avenue). Finally, another possibility of future work could be to study which visualization would be appropriate for showing the relationships between the administrative registers and the statistic operations that are linked (because this register produces this operation) at the same time that the visualization shows the clusters of nodes by thematic.